\documentclass[12pt,a4paper]{iopart}

\usepackage{graphicx,subfig}

\newcommand{\caco}{CuAl$_{1-x}$Cr$_x$O$_2$}
\newcommand{\cao}{CuAlO$_2$}
\newcommand{\cco}{CuCrO$_2$}
\newcommand{\cfo}{CuFeO$_2$}

\begin{document}

\title[Delafossite solid solution (\cao)$_{1-x}$(\cco)$_x$]{Structural and magnetic characterization of the complete delafossite solid solution (\cao)$_{1-x}$(\cco)$_x$} 

\author{Phillip T. Barton, Ram Seshadri} \eads{pbarton@mrl.ucsb.edu, seshadri@mrl.ucsb.edu}
\address{Materials Department and Materials Research Laboratory\\
University of California, Santa Barbara, CA, 93106, USA}

\author{Andrea Kn\"{o}ller}
\address{Institut f\"{u}r Materialwissenschaft,\\
Universit\"{a}t Stuttgart 70569, Germany}

\author{Matthew J. Rosseinsky} \eads{M.J.Rosseinsky@liverpool.ac.uk}
\address{Department of Chemistry, \\
University of Liverpool, L69 7ZD, UK}

\date{\today}

\begin{abstract}
We have prepared the complete delafossite solid solution series between diamagnetic \cao\/ and the $t_{2g}^3$ frustrated antiferromagnet \cco\/. The evolution with composition $x$ in \caco\/ of the crystal structure and magnetic properties has been studied and is reported here. The room-temperature unit cell parameters follow the V\'egard law and increase with $x$ as expected. The $\mu_{\rm{eff}}$ is equal to the Cr$^{3+}$ spin-only $S = 3/2$ value throughout the entire solid solution. $\Theta_{\rm{CW}}$ is negative, indicating that the dominant interactions are
antiferromagnetic, and its magnitude increases with Cr substitution. For dilute Cr compositions, the nearest-neighbor exchange coupling constant $J_{\rm{BB}}$ was estimated by mean-field theory to be 3.0\,meV. Despite the sizable $\Theta_{\rm{CW}}$, long-range antiferromagnetic order does not develop until $x$ is almost 1, and is preceeded by glassy behavior. Data presented here, and that on dilute Al-substitution from Okuda \textit{et al.}, suggest that the reduction in magnetic frustration due to the presence of non-magnetic Al does not have as dominant an effect on magnetism as chemical disorder and dilution of the magnetic exchange. For all samples, the 5\,K isothermal magnetization does not saturate in fields up to 5\,T and minimal hysteresis is observed. The presence of antiferromagnetic interactions is clearly evident in the sub-Brillouin behavior with a reduced magnetization per Cr atom. An inspection of the scaled Curie plot reveals that significant short-range antiferromagnetic interactions occur in \cco\/ above its N\'eel temperature, consistent with its magnetic frustration. Uncompensated short-range behavior is present in the Al-substituted samples and is likely a result of chemical disorder.
\end{abstract}
\maketitle

\section{Introduction}

\cao\/ and \cco\/ are $p$-type transparent conducting oxides (TCO) that are of significant interest for their intrinsic $p$-type behavior.\cite{Kawazoe_Nature1997,Nagarajan_JAP2001} \cco\/ is also a $S$ = 3/2 Heisenberg triangular lattice antiferromagnet (TLA), which makes it a promising candidate for studying magnetic frustration. Recent study of \cco\/ is spurred by its multiferroic behavior which arises from its spiral spin order,\cite{Tokura_AM2010} and has been extensively investigated by Kimura \textit{et al.}\cite{Kimura_PRL2009}

Neutron diffraction studies have been essential in explaining the magnetism and multiferroic behavior of \cco. The first neutron study by Kadowaki \textit{et al.} revealed that \cco\/ has an antiferromagnetic out-of-plane 120$^{\circ}$ spin structure and short correlation length along the $c$ axis.\cite{Kadowaki_JPCM1990} Further study by Poienar \textit{et al.} narrowed the magnetic structure possibilities to either helicoidal or cycloidal, and investigated the effect of Mg substitution.\cite{Poienar_PRB2009} Soda \textit{et al.} confirmed a noncollinear helicoidal magnetic structure through triple-axis spin-polarized neutron scattering experiments on a single crystal.\cite{Soda_JPSJ2009} Such a magnetic structure also occurs for \cfo\/ under an applied magnetic field or with Al substitution, and was found to give rise to ferroelectricity.\cite{Kimura_PRB2006} This is consistent with a theoretical model proposed by Arima, which shows that a noncollinear helical spin structure and spin-orbit coupling give rise to the multiferroic behavior.\cite{Arima_JPSJ2007} Similar to other TLAs, the presence of two
magnetic transitions in \cco\/ was revealed by a careful further examination of a single crystal.\cite{Kimura_PRB2008} Additional study of \cco\/ included inelastic neutron scattering to map out the spin dynamics of the system.\cite{Poienar_PRB2010} The results are consistent with the work of Kimura \textit{et al.},\cite{Kimura_PRL2009} and reinforce the critical role
of next-nearest-neighbor exchange interactions in stabilizing magnetic order.

The delafossite structure has also been of interest because it hosts both metallic and insulating behavior, as is well exemplified by the metal-insulator transition in a partial solid solution between AgNiO$_2$ and AgCoO$_2$.\cite{Shin_SSC1993} To investigate the nature of such behavior, the electronic structures of many delafossites have been investigated by density
functional theory (DFT) calculations, and particular interest has been paid to the Cu-containing $p$-type TCOs. Evidence has been shown that these derive their $p$-type conductivity from Cu vacancies that form because of the easy oxidation of Cu$^{1+}$ to Cu$^{2+}$.\cite{Scanlon_JMC2011} Additionally, the extent of M-M bonding in delafossites has been examined with DFT for a multitude of different A and B cations.\cite{Seshadri_CM1998,Kandpal_SSC2002} In these studies the authors note that both the A and B site cations contribute to the electronic structure near the Fermi level. They also found that the A site $d_{z^2}$ orbitals are responsible for the highly disperse bands and may be important for stabilizing metallic ground states. Relevant
to the systems of interest here, Scanlon \textit{et al.} studied the effect of Cr substitution on the electronic structure of \cao.\cite{Scanlon_PRB2009}

The effect of spin dilution and doping on the magnetism of \cco\/ have been extensively investigated by Okuda \textit{et. al} through magnetometry, electrical transport,\cite{Okuda_PRB2005} heat capacity measurements,\cite{Okuda_PRB2008} and neutron diffraction.\cite{Okuda_JPSJ2011} Mg substitution sharpens the antiferromagnetic transition and thus demonstrates that hole carriers are relevant to the magnetic ground state of \cco\/.\cite{Okuda_PRB2005} Al substitution blurs the transition, causes the evolution of spin-glass behavior, and causes a crossover from 3D to 2D magnons as evidenced by heat capacity data.\cite{Okuda_JPSJ2011} It also causes a gradual suppression of the \cco\/ magnetic peaks in neutron diffraction patterns.

In this contribution, we investigate the complete delafossite solid solution between diamagnetic \cao\/ and the $t_{2g}^3$ frustrated antiferromagnet \cco. While many chemical substitutions have been performed on both end-member compounds, this is the first time a complete solid solution has been prepared. This work follows a recent study of a perovskite solid solution,\cite{Barton_PRB2011} where we used magnetic dilution to probe the ferromagnet SrRuO$_3$. This study is also guided by previous work on magnetic frustration in MCr$_2$O$_4$ spinels, where Cr sits on the pyrochlore B-sublattice.\cite{Melot_JPCM2009,Shoemaker_PRB2010} In \caco, the addition of Cr introduces localized spins that are randomly distributed on the B site. These spins begin interacting as their concentration is increased, and frustration arises due to the triangular topology of the delafossite crystal structure. Powder x-ray diffraction and magnetometry results on \caco\/ support that a well-behaved solid solution is formed. Throughout the solid solution, $\mu_{\rm{eff}}$ is equal to the Cr$^{3+}$ spin-only $S = 3/2$ value. The strength of the mean-field antiferromagnetic interactions markedly increases with Cr content, though glassy long-range order does not occur until $x \approx 0.75$. Magnetic saturation does not occur in isothermal magnetization sweeps, and antiferromagnetic interactions are evident by the sub-Brillouin behavior of all samples. A scaled Curie plot reveals the presence of short-range interactions that occur due to frustration and chemical disorder.

\section{Experimental details}

Polycrystalline \caco\/ pellets were prepared using solid-state reactions at high temperatures. Stoichiometric amounts of Cu$_2$O, Al$_2$O$_3$, and Cr$_2$O$_3$ powders were ground with an agate mortar and pestle, pressed at 100\,MPa, and heated in air to temperature for 24\,h, and again for 48\,h with an intermediate grinding, and then allowed to cool to room temperature. The furnace was heated and cooled at a rate of 2\,$^{\circ}$C/min, and in accordance with previous preparations, firing temperatures were 1000\,$^{\circ}$C for \cco\/\cite{Doumerc_MRB1986}, 1200\,$^{\circ}$C for \cao\cite{Ishiguro_JSSC1981}, and between the two for intermediate compositions. The pellets were placed on beds of powders of the same composition to avoid reaction with the Al$_2$O$_3$ crucible. Structural characterization by room-temperature laboratory x-ray diffraction was performed on a Philips X'Pert diffractometer with Cu-$K_\alpha$ radiation. Select samples were also examined by high resolution powder synchrotron x-ray diffraction at the 11-BM beamline at the Advanced Photon Source, Argonne National Laboratory. Rietveld\cite{Rietveld_JAC1969} refinement was performed using the XND Rietveld code.\cite{Berar_XND_1998} Crystal structures were visualized using VESTA.\cite{Momma_VESTA_2008} Magnetization was measured using a Quantum Design MPMS 5XL SQUID magnetometer.

\section{Results and Discussion} 

\subsection{Structure}

\begin{figure}
\centering
\subfloat{\includegraphics[width=0.9in]{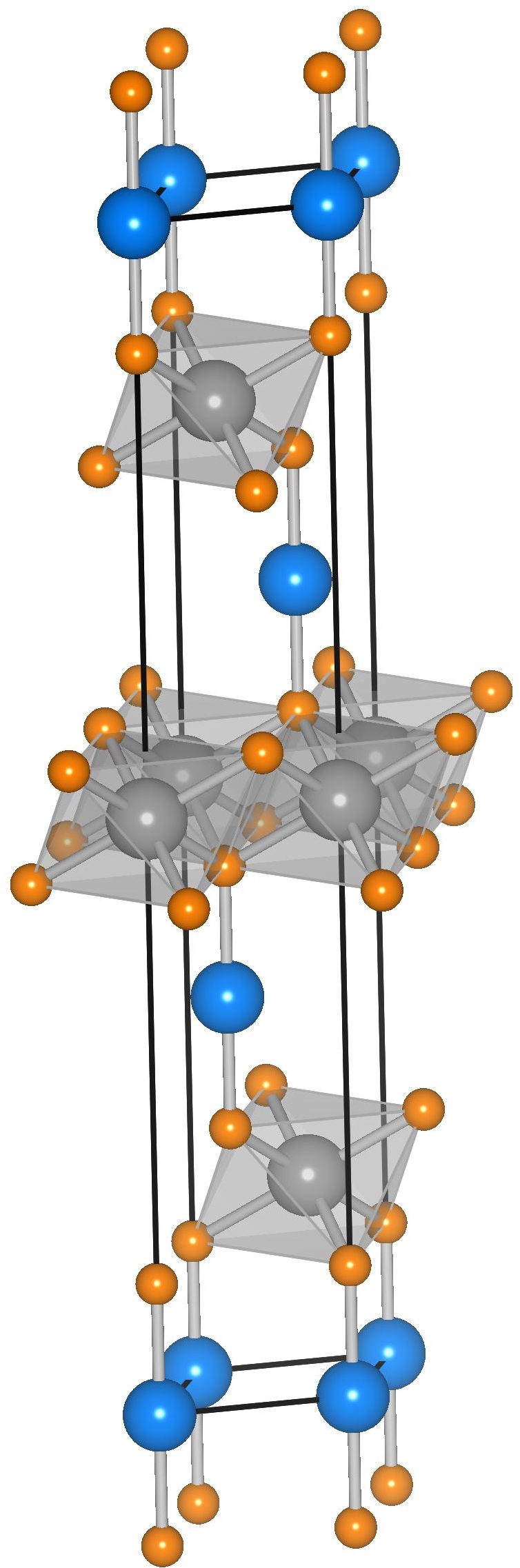}}\hspace{0.15in}
\subfloat{\includegraphics[width=3in]{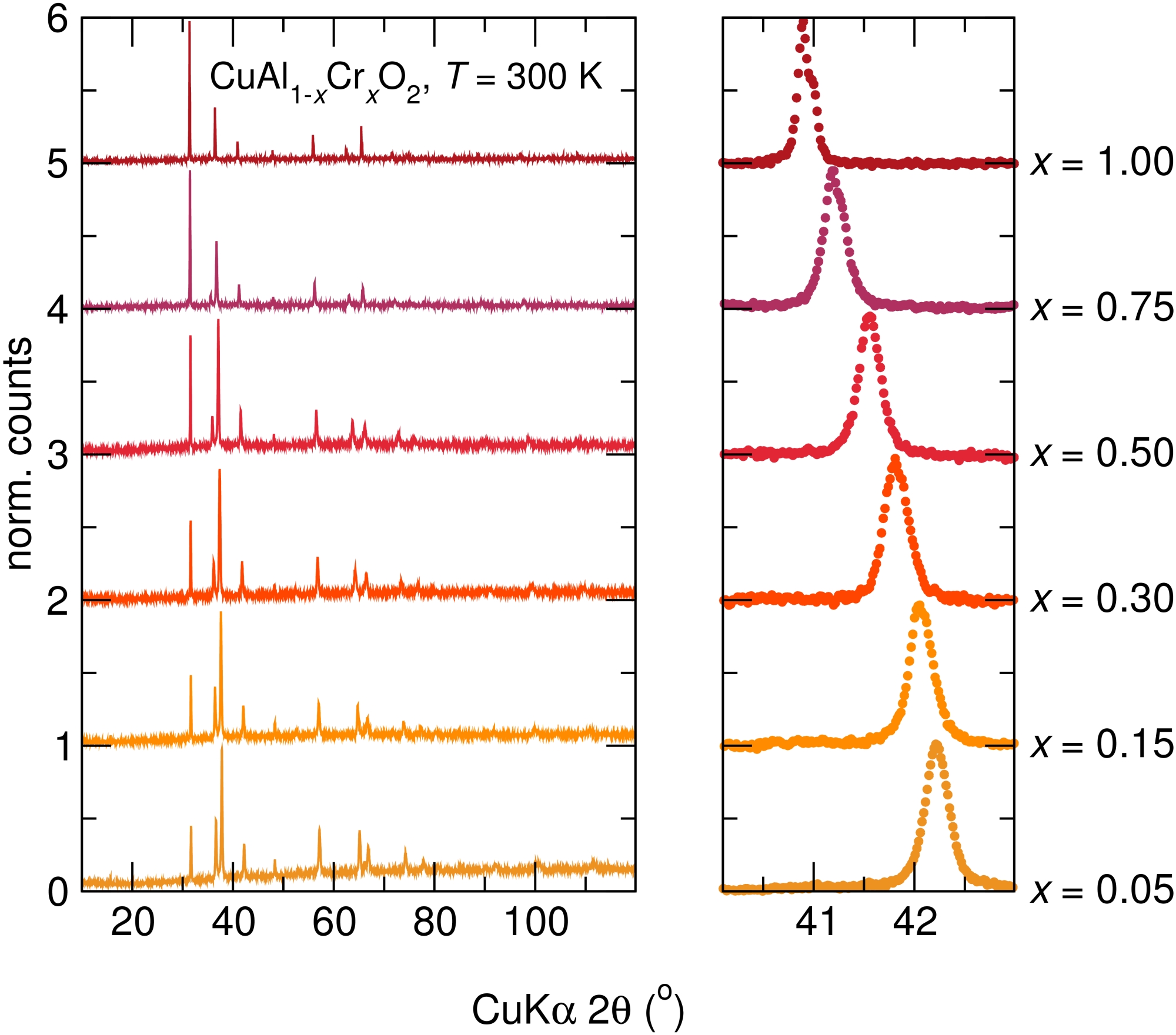}}
\caption{(Color online) The left panel displays the ABO$_2$ delafossite crystal structure with A in blue, B in grey, and O in orange. The middle panel shows room-temperature laboratory powder x-ray diffraction data for \caco. The right panel shows a close-up of the (10$\bar{4}$) peak as it evolves across the solid solution.}
\label{fig:xrd}
\end{figure}

\begin{figure}
\centering
\includegraphics[width=3in]{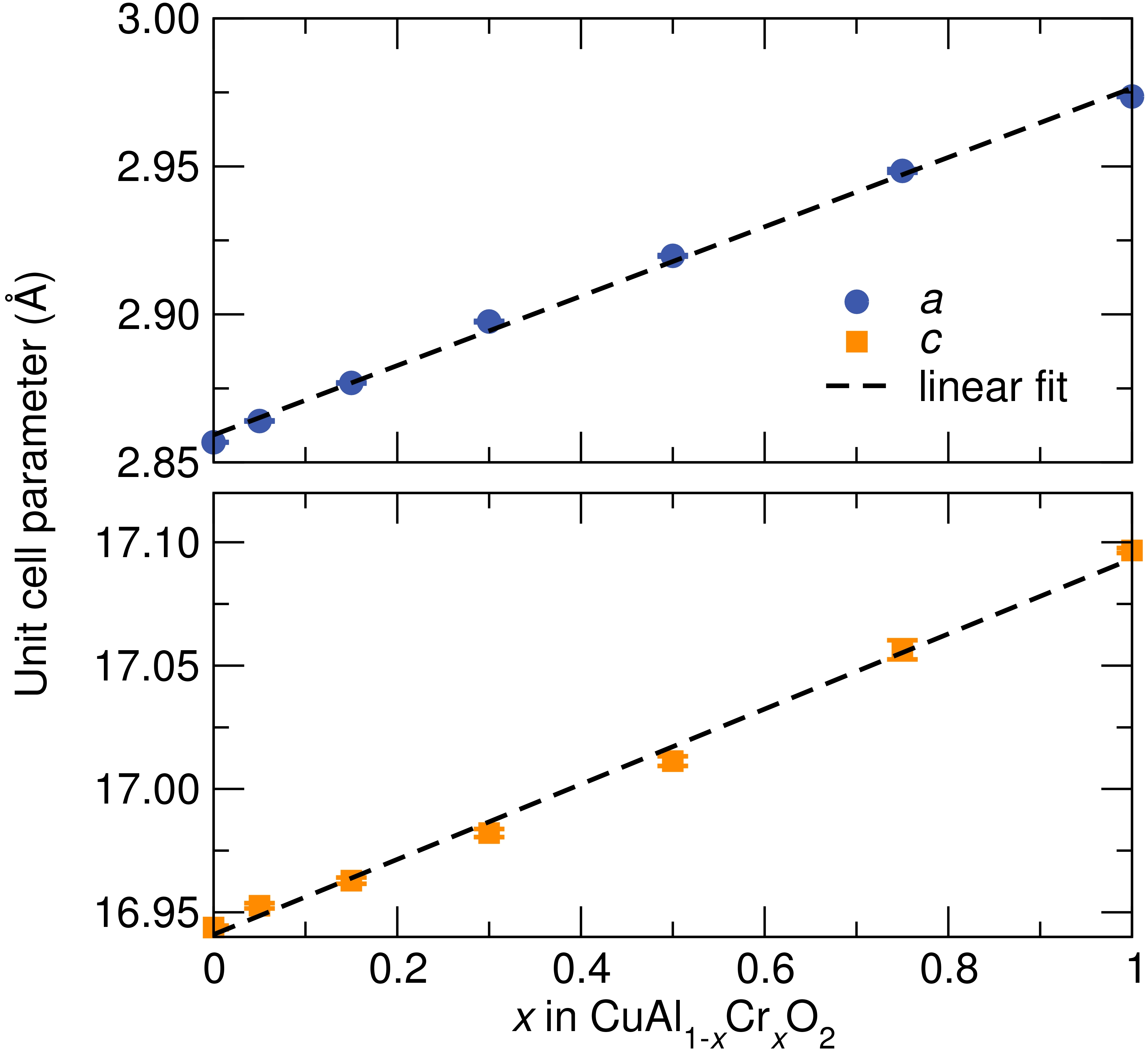}\\
\caption{(Color online) Unit cell parameters $a$ and $c$ of the $3R$ delafossite crystal structure for \caco\/ as determined by Rietveld refinement. Linear fits to the data demonstrate that the V\'egard law is followed. Error bars are included, but are smaller than the symbol size for most data points.}
\label{fig:struct_params}
\end{figure}

\begin{table*}
\caption{Unit cell parameters and cell volume for \caco\, obtained from Rietveld refinement of laboratory powder x-ray diffraction data in space group $R\bar{3}m$ (No. 166). Cu sits at (0,0,0), (Al/Cr) at (0,0,0.5) and O at (0,0,$z$).}
\label{tab:struc}
\centering
\begin{tabular}{lccccccccc}
\hline
\hline
$x$ &0.00 &0.05 &0.15 &0.30 &0.50 &0.75 &1.00\\
\hline
$a$ (\AA) &2.856(8) &2.864(0) &2.876(8) &2.897(6) &2.919(8) &2.948(5) &2.973(6)\ \\
$c$ (\AA) &16.94(4) &16.95(3) &16.96(3) &16.98(2) &17.01(1) &17.05(6) &17.09(7)\ \\
$V$ (\AA$^3$) &138.2(8) &139.0(5) &140.3(9) &142.5(8) &145.0(2) &148.2(8) &151.1(8)\ \\
\hline
\hline
\end{tabular}
\end{table*}

Room-temperature laboratory and synchrotron powder x-ray diffraction demonstrate that the entire solid solution between \cao\/ and \cco\/ can be assigned to the rhombohedral $3R$ delafossite crystal structure, space group $R\bar{3}m$ (No. 166). All of the observed Bragg peaks in laboratory data are accounted for by the $3R$ structure and support the single phase nature of the samples (Figure~\ref{fig:xrd}). The refined unit cell parameters and volume are tabulated in Table~\ref{tab:struc} and the composition dependence is displayed graphically in Figure~\ref{fig:struct_params}. Upon substituting Cr into \cao, the $a$ and $c$ parameters increase, as expected based on the Shannon-Prewitt effective ionic radii.\cite{Shannon_ACB1969} Their linear dependence on composition is consistent with the V\'egard law and supports that a true solid solution is formed. Select samples ($x$ = 0.05, 0.75, and 1.00) were further characterized by high-resolution powder synchrotron x-ray diffraction, which revealed the presence of small amounts of CuO and Cr$_2$O$_3$ (not shown). The thorough characterization by x-ray diffraction provides strong evidence for the structure and composition of the delafossite solid solution series \caco.

\subsection{Magnetism}

\begin{table*}
\caption{Magnetic characteristics of \caco. $\Theta_{CW}$ and $\mu_{\rm{eff}}$ were obtained by Curie-Weiss analysis. $M$ is the magnetization at 5\,K and 5\,T, though none of the samples reach saturation under these conditions.}
\label{tab:mag}
\centering
\begin{tabular}{lcccccccccccc}
\hline
\hline
$x$ &0.05 &0.15 &0.30 &0.50 &0.75 &1.00\\
\hline
$\Theta_{CW}$ (K) &$-$24.7 &$-$74.0 &$-$155 &$-$209 &$-$202 &$-$158 \\
$\mu_{\rm{eff}}$ ($\mu_B$/Cr) &3.92 &3.78 &3.98 &3.98 &3.80 &3.60 \\
$M$ ($\mu_B$/Cr) &1.16 &0.516 &0.367 &0.202 &0.0381 &0.0170 \\
\hline
\hline
\end{tabular}
\end{table*}

\begin{figure}
\centering\includegraphics[width=3in]{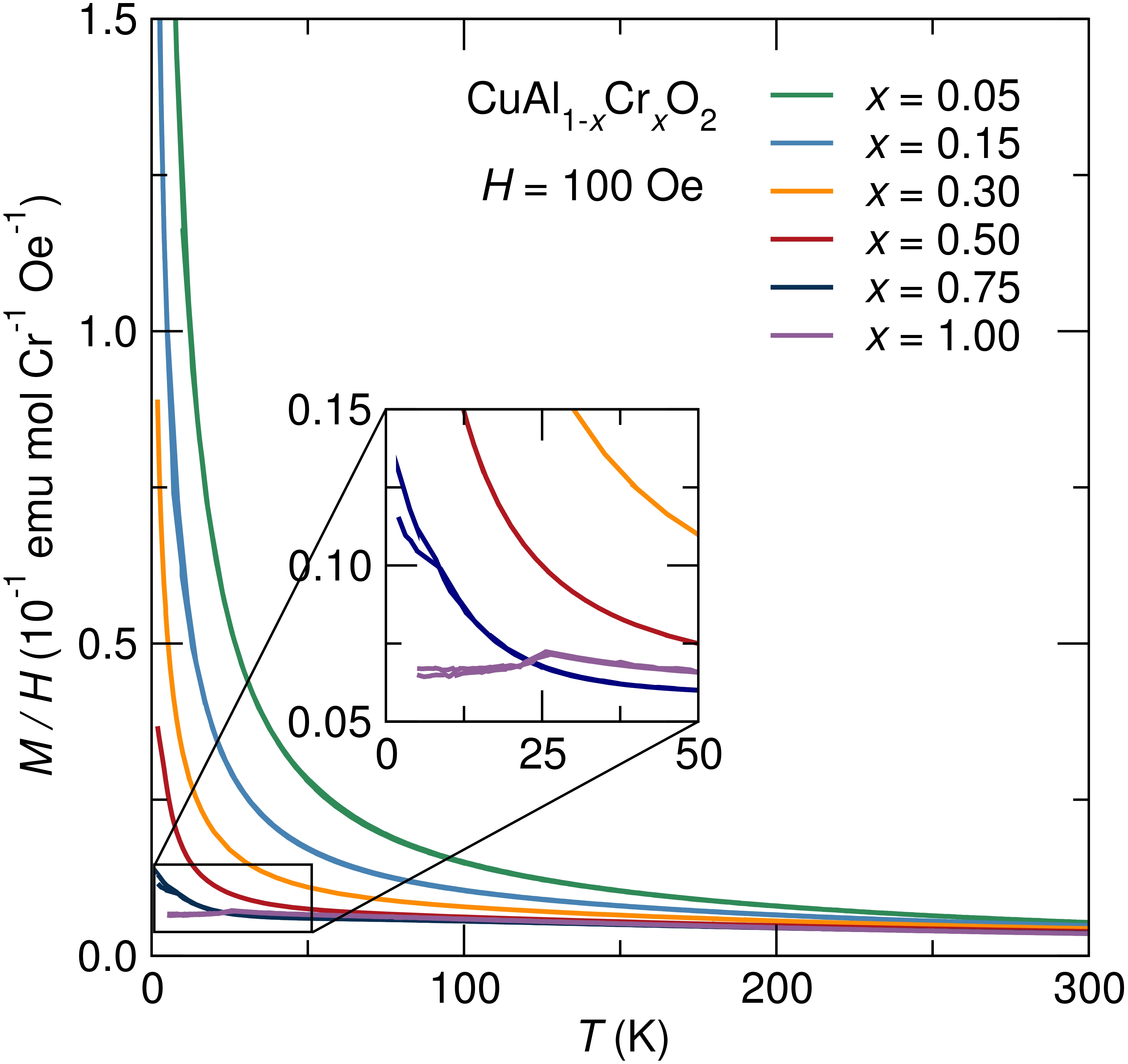}\\
\caption{(Color online) Zero-field cooled and field-cooled magnetic susceptibility as collected under a DC magnetic field of 100\,Oe.}
\label{fig:M-T}
\end{figure}

\begin{figure}
\centering
\includegraphics[width=3in]{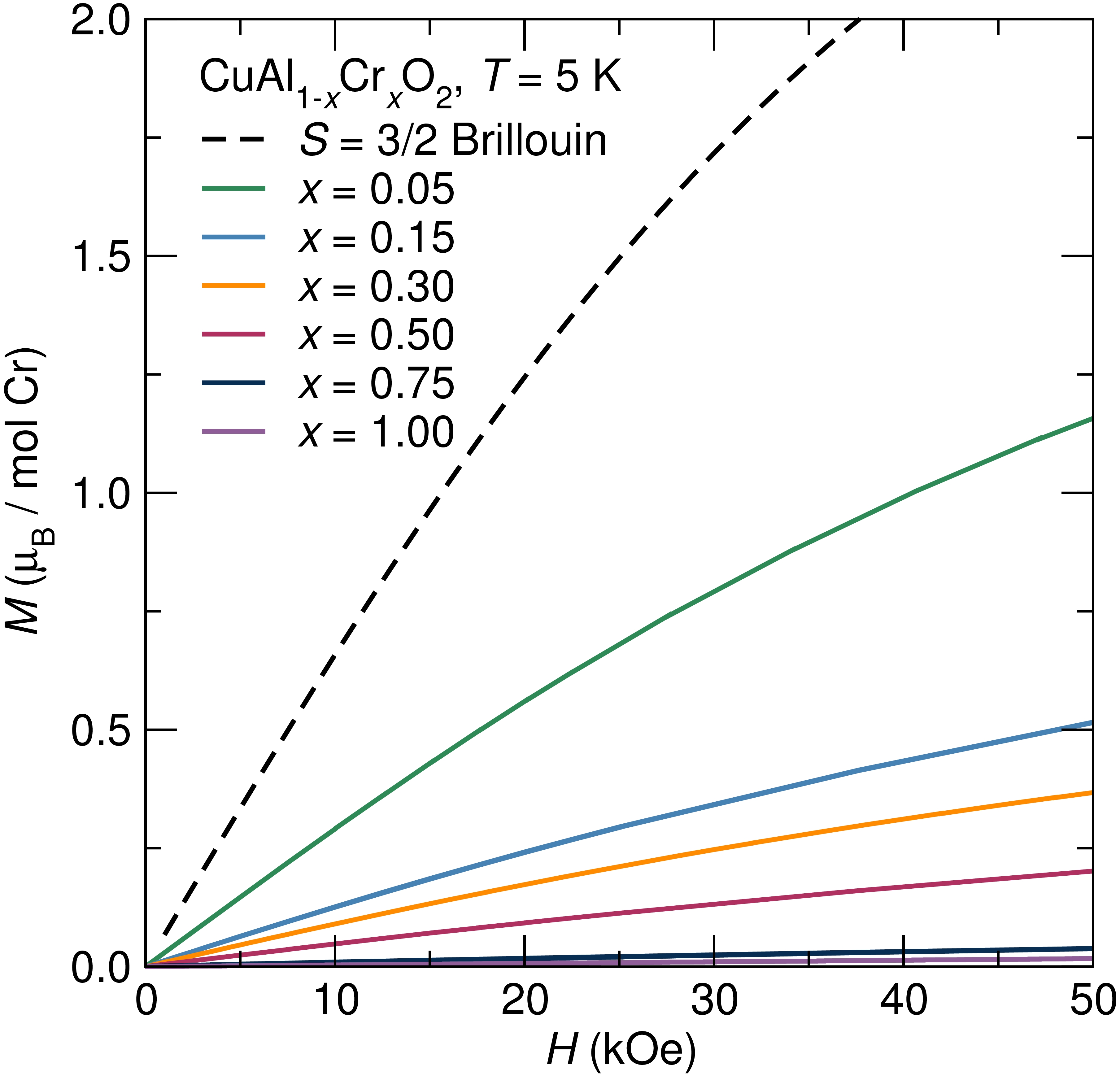}\\
\caption{(Color online) Magnetization as a function of applied DC magnetic field at 5\,K. Data were acquired in a loop from 0\,T, to 5\,T, to $-$5\,T, and back to 0\,T, though only the first quadrant is shown as no significant hysteresis was observed.}
\label{fig:M-H}
\end{figure}

\begin{figure}
\centering\includegraphics[width=3in]{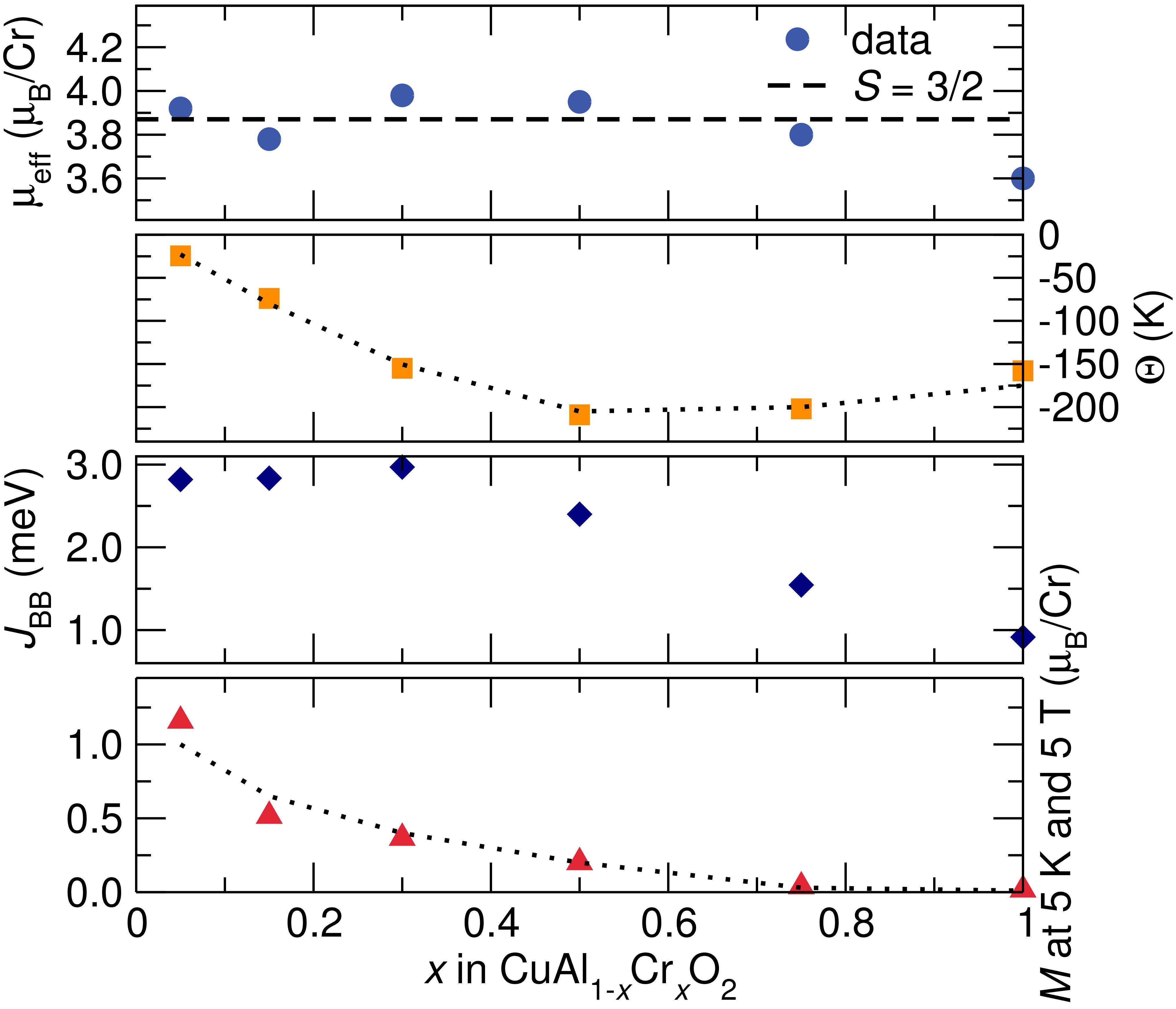}\\
\caption{(Color online) Magnetic characteristics (a) $\mu_{\rm{eff}}$, (b) $\Theta_{\rm{CW}}$, (c) $J_{\rm{BB}}$, and (d) $M$(2\,K, 5\,T) as a function of composition. In (a), the dashed horizontal line indicates the expected $S$-only value for Cr$^{3+}$. The dotted lines in the other panels are guides to the eye.}
\label{fig:mag_prop}
\end{figure}

In \caco, magnetism evolves as we alloy a diamagnet with an antiferromagnet, and is complicated by the geometrically-induced magnetic frustration of the delafossite crystal structure. Important magnetic characteristics of the solid solution, including Curie-Weiss effective moment $\mu_{\rm{eff}}$, theoretical magnetic ordering temperature $\Theta_{CW}$, B site nearest-neighbor magnetic exchange $J_{\rm{BB}}$, and $M$ at 5\,K and 5\,T, are summarized graphically in Figure~\ref{fig:mag_prop} and are tabulated in Table~\ref{tab:mag}. Zero-field cooled (ZFC) and field-cooled (FC) magnetization data were collected between 2\,K and 380\,K under a 100\,Oe DC magnetic field and are presented in Figure~\ref{fig:M-T}. We observe an antiferromagnetic cusp in the magnetization of \cco\/ at 25\,K, in accord with previous characterization.\cite{Doumerc_MRB1986} From Curie-Weiss analysis, we find that, as expected for octahedral Cr$^{3+}$, $\mu_{\rm{eff}}$ is nearly equal to the spin-only $S=3/2$ value of 3.82 $\mu_{B}$ throughout the solid solution. As evidenced by the negative $\Theta_{CW}$, the dominant magnetic interactions are antiferromagnetic. The magnitude of $\Theta_{CW}$ is small for samples with low Cr content, but it quickly increases with $x$ as Cr-O-Cr superexchange interactions become more prevalent. The nearest-neighbor magnetic exchange coupling $J_{\rm{BB}}$ can be estimated using a mean-field Heisenberg  model according to  $J = 3\Theta k_{\rm{B}}/[2Z_{\rm{eff}}S(S+1)]$, where $Z_{\rm{eff}} = 6x$ is the number of nearest-neighbor interactions. Such analysis reveals that $J_{\rm{BB}}$ is close to 3.0\,meV for dilute Cr compositions and decreases with increasing $x$. This decrease is a result of the mean-field approximation breaking down as Cr composition increases and next-nearest-neighbor interactions become important. The magnitude of $J_{\rm{BB}}$ is similar to  previous studies of Cr-O-Cr exchange in structurally similar spinels with edge-sharing octahedra of Cr.\cite{Melot_JPCM2009,Ederer_PRB2007,Lawes_PRB2006} A zfc-fc splitting occurs in the $x$ = 0.75 sample below 8\,K, which is consistent with splittings seen in low Al content samples studied by Okuda \textit{et al.}\cite{Okuda_JPSJ2011} Such behavior is attributed to chemical disorder that results in spin-glass behavior. The dependence of magnetization $M$ on field for \caco\/ at 5\,K is displayed in Figure~\ref{fig:M-H}. None of the samples reach magnetic saturation. All of the $M-H$ traces are sub-Brillouin, indicating significant antiferromagnetic exchange which increases in magnitude with $x$. Hysteresis is only observed in the glassy $x$ = 0.75 sample, but with a small $H_C$ value of 200\,Oe.

Curie-Weiss analysis was performed in the high-temperature region where the inverse magnetization was linear, which was dependent on sample, but typically 250\,K to 380\,K. For the very dilute $x = 0.05$ sample, a temperature independent term was included in the Curie-Weiss analysis to capture the diamagnetism of the sample. Without this term, the fit was significantly poorer and the results were not consistent with the solid solution, with an extracted $\mu_{\rm{eff}}$ and $\Theta_{CW}$ of approximately 3.5\,$\mu_{B}$ and $+$5.6\,K.

\begin{figure}
\centering
\includegraphics[width=3in]{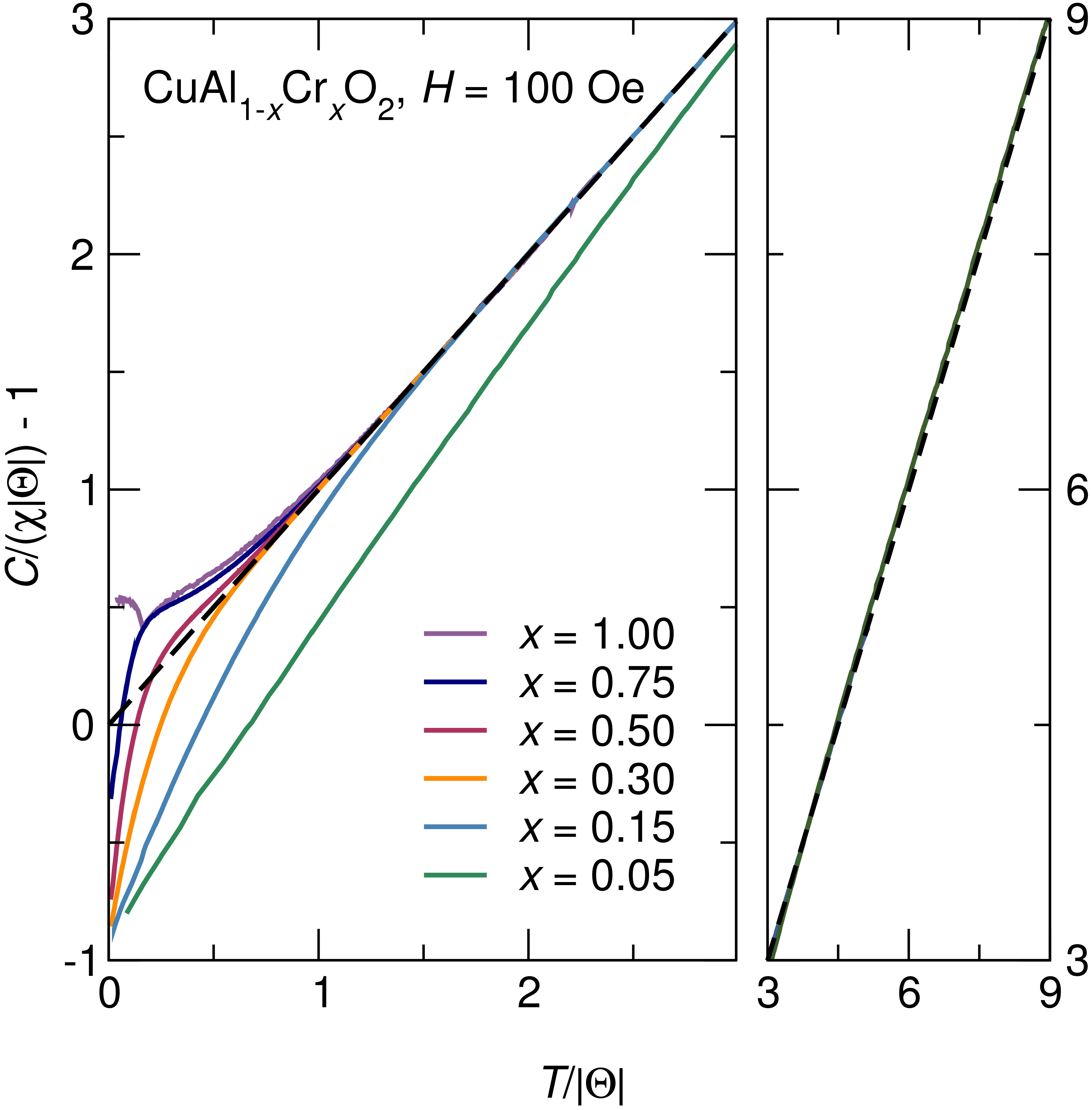}\\
\caption{(Color online) Scaled inverse magnetic susceptibility as a function of scaled temperature, as described by equation \ref{eqn:CW}. The dashed black line represents ideal Curie-Weiss paramagnetism, and deviations from it correspond to short-range interactions. The right panel demonstrates that all samples follow ideal Curie-Weiss behavior at high temperatures.}
\label{fig:curie}
\end{figure}

The Curie-Weiss relation $\chi=C/(T-\Theta_{CW})$ can be recast according to: 

\begin{equation}
\frac{C}{\chi|\Theta_{CW}|}+\mbox{sgn}(\Theta_{CW})=\frac{T}{|\Theta_{CW}|}
\label{eqn:CW}
\end{equation}

\noindent
which allows normalization of susceptibility-temperature plots as shown in Figure~\ref{fig:curie}. The utility of such plots has been amply demonstrated in the analysis of other solid-solution systems.\cite{Melot_JPCM2009} At temperatures above the long-range ordering temperature, positive deviations from the ideal Curie-Weiss line reflect the presence of compensated antiferromagnetic short-range interactions, while negative deviations reflect uncompensated behavior (ferromagnetism or ferrimagnetism). Consistent with its magnetic frustration, \cco\/ displays short-range antiferromagnetic interactions well above its N\'eel temperature. While Curie-Weiss analysis shows that the dominant long-range interactions in \caco\/ are antiferromagnetic, the scaled Curie plot reveals the presence of short-range uncompensated behavior in all of the Al-substituted samples which is likely a result of chemical disorder. For the small $x$ samples, their behavior follows the ideal Curie-Weiss until low temperatures at which they deviate below the line. The $x$ = 0.75 sample displays a more complex behavior, however, as upon cooling it first displays short-range antiferromagnetic interactions, similar to \cco\/, and only at low temperatures does uncompensated behavior emerge. When taken in combination with the slight zfc-fc splitting and small $H_C$, this behavior may be consistent with a tendency towards an antiferromagnetic ground state that is disrupted by the freezing of spins into a spin-glass where uncompensated moments are present.

\section{Conclusion}

We prepared a complete delafossite solid solution between diamagnetic \cao\/ and the $t_{2g}^3$ frustrated antiferromagnet \cco\/ and characterized its structural and magnetic properties. All observed laboratory x-ray diffraction peaks correspond to the delafossite $3R$ crystal structure.  A true solid solution forms, as evidenced by the adherence of unit cell parameters to the V\'egard law and by the magnetic behavior. $\mu_{\rm{eff}}$ is equal to the Cr$^{3+}$ spin-only $S = 3/2$ value throughout the solid solution, while $\Theta_{CW}$ is negative and its magnitude increases with $x$. $J_{\rm{BB}}$ was estimated by mean-field theory to be 3.0\,meV for dilute Cr compositions. Magnetic saturation does not occur at 5\,K and 5\,T, and the sub-Brillouin behavior is consistent with strong antiferromagnetic interactions. Inspection of a scaled inverse magnetic susceptibility plot reveals that significant short-range antiferromagnetic interactions exist in \cco\/ above $T_N$, while uncompensated short-range behavior are present in the Al-substituted samples. These observations can be explained by magnetic frustration and chemical disorder. The results are relevant for understanding magnetic frustration and for the tuning of physical properties through chemical substitution.

\section{Acknowledgements}

We acknowledge support through a Materials World Network Award from the NSF (DMR 0909180) in Santa Barbara, and from the EPSRC (EP/G065314/1) in Liverpool. PTB is supported by the NSF Graduate Research Fellowship Program. AK is supported by the IMI Program of the National Science Foundation under Award No. DMR 0843934, and the UCSB-MPG Program for International Exchange in Materials Science. The research carried out here made extensive use of shared experimental facilities of the Materials Research Laboratory: an NSF MRSEC, supported by NSF DMR 1121053. The MRL is a member of the the NSF-supported Materials Research Facilities Network (www.mrfn.org). Use of data from the 11-BM beamline at the Advanced Photon Source was supported by the U.S. Department of Energy, Office of Science, Office of Basic Energy Sciences, under Contract No. DE-AC02-06CH11357.

\clearpage


\clearpage

\begin{thebibliography}{10}

\bibitem{Kawazoe_Nature1997}
H. Kawazoe, M. Yasukawa, H. Hyodo, M. Kurita, H. Yanagi, and H. Hosono, \textit{Nature} \textbf{389}, 939 (1997).

\bibitem{Nagarajan_JAP2001}
R. Nagarajan, A. D. Draeseke, A. W. Sleight, and J. Tate, \textit{J. App. Phys.} \textbf{89}, 8022 (2001).

\bibitem{Tokura_AM2010}
Y. Tokura and S. Seki, \textit{Adv. Mater.} \textbf{22}, 1554 (2010).

\bibitem{Kimura_PRL2009}
K. Kimura, H. Nakamura, S. Kimura, M. Hagiwara, and T. Kimura, \textit{Phys. Rev. Lett.} \textbf{103} 107201 (2009).

\bibitem{Kadowaki_JPCM1990}
H. Kadowaki, H. Kikuchi, and Y. Ajiro, \textit{J. Phys.: Cond. Mat.} \textbf{2}, 4485 (1990).

\bibitem{Poienar_PRB2009}
M. Poienar, F. Damay, C. Martin, V. Hardy, A. Maignan, and G. Andr\'e, \textit{Phys. Rev. B} \textbf{79}, 014412 (2009).

\bibitem{Soda_JPSJ2009}
M. Soda, K. Kimura, T. Kimura, M. Matsuura, and K. Hirota, \textit{J. Phys. Soc. Jap.} \textbf{78} 124703 (2009).

\bibitem{Kimura_PRB2006}
T. Kimura, J. C. Lashley, and A. P. Ramirez, \textit{Phys. Rev. B} \textbf{73}, 220401 (2006).

\bibitem{Arima_JPSJ2007}
T. Arima, \textit{J. Phys. Soc. Jap.} \textbf{76}, 073702, (2007).

\bibitem{Kimura_PRB2008}
K. Kimura, H. Nakamura, K. Ohgushi, and T. Kimura, \textit{Phys. Rev. B} \textbf{78}, 140401 (2008).

\bibitem{Poienar_PRB2010}
M. Poienar, F. Damay, C. Martin, J. Robert, and S. Petit, \textit{Phys. Rev. B} \textbf{81} 104411 (2010).

\bibitem{Shin_SSC1993}
Y. J. Shin, J. P. Doumerc, P. Dordor, M. Pouchard, P. J. Hagenmuller, \textit{Solid State Chem.} \textbf{107}, 194 (1993).

\bibitem{Scanlon_JMC2011}
D. O. Scanlon and G. W. Watson, \textit{J. Mater. Chem.} \textbf{21}, 3655 (2011).

\bibitem{Seshadri_CM1998}
R. Seshadri, C. Felser, K. Thieme, and W. Tremel, \textit{Chem. Mater.} \textbf{10}, 2189 (1998).

\bibitem{Kandpal_SSC2002}
H. C. Kandpal and R. Seshadri, \textit{Solid State Sci.} \textbf{4}, 1045 (2002).

\bibitem{Scanlon_PRB2009}
D. O. Scanlon, A. Walsh, B. J. Morgan, G. W. Watson, D. J. Payne, and R. G. Egdell, \textit{Phys. Rev. B} \textbf{79}, 035101 (2009).

\bibitem{Okuda_PRB2005}
T. Okuda, N. Jufuku, S. Hidaka, and N. Terada, \textit{Phys. Rev. B} \textbf{72}, 144403 (2005).

\bibitem{Okuda_PRB2008}
T. Okuda, Y. Beppu, Y. Fujii, T. Onoe, N. Terada, and S. Miyasaka, \textit{Phys. Rev. B} \textbf{77}, 134423 (2008).

\bibitem{Okuda_JPSJ2011}
T. Okuda, K. Uto, S. Seki, Y. Onose, Y. Tokura, R. Kajimoto, and Masaaki Matsuda, \textit{J. Phys. Soc. Jap.} \textbf{80}, 014711 (2011).

\bibitem{Barton_PRB2011}
P. T. Barton, R. Seshadri, and M. J. Rosseinsky, \textit{Phys. Rev. B} \textbf{83}, 064417 (2011).

\bibitem{Melot_JPCM2009}
B. C. Melot, J. E. Drewes, R. Seshadri, E. M. Stoudenmire, and A. P. Ramirez, \textit{J. Phys.: Cond. Matter} \textbf{21}, 216007 (2009).

\bibitem{Shoemaker_PRB2010}
D. P. Shoemaker and R. Seshadri, \textit{Phys. Rev. B} \textbf{82}, 214107 (2010).

\bibitem{Ishiguro_JSSC1981}
T. Ishiguro, A. Kitazawa, N. Mizutani, and M. Kato, \textit{J. Sol. State Chem.} \textbf{40}, 170 (1981).

\bibitem{Doumerc_MRB1986}
J.-P. Doumerc, A. Wichainchai, A. Ammar, M. Pouchard, and P. Hagenmuller, \textit{Mater. Res. Bull.} \textbf{21}, 745 (1986).

\bibitem{Rietveld_JAC1969}
H. M. Rietveld, \textit{J. Appl. Crystallogr.} \textbf{2}, 65 (1969).

\bibitem{Berar_XND_1998}
J. B\'erar and G. Baldinozzi, \textit{IUCr-CPD Newsletter} \textbf{20}, 3 (1998).

\bibitem{Momma_VESTA_2008}
K. Momma and F. Izumi, \textit{J. Appl. Crystallogr.} \textbf{41}, 653 (2008).

\bibitem{Shannon_ACB1969}
R. D. Shannon and C. T. Prewitt, \textit{Acta Crystallogr. B} \textbf{25}, 925 (1969).

\bibitem{Ederer_PRB2007}
C. Ederer and M. Komelj, \textit{Phys. Rev. B} \textbf{76}, 064409 (2007).

\bibitem{Lawes_PRB2006}
G. Lawes, B. Melot, K. Page, C. Ederer, M. A. Hayward, Th. Proffen, and R. Seshadri, \textit{Phys. Rev. B} \textbf{74}, 024413 (2006).

\end{thebibliography}
\end{document}